**Digital Twin-Based Intrusion Detection for Vehicle Powertrain CAN Bus Systems**

**Araf Rahman***
Glenn Department of Civil Engineering
Clemson University, Clemson, South Carolina, 29634
Email: arafr@clemson.edu

**M Sabbir Salek, Ph.D.**
Glenn Department of Civil Engineering
Clemson University, Clemson, South Carolina, 29634
Email: msalek@clemson.edu

**Mashrur Chowdhury, Ph.D., P.E.**
Glenn Department of Civil Engineering
Clemson University, Clemson, South Carolina, 29634
Email: mac@clemson.edu

*Corresponding Author

## ABSTRACT

**Objectives:** Existing automotive intrusion detection systems (IDSs) for the Controller Area Network (CAN) largely target discrepancies in message timing, frequency, or sequencing, and cannot detect attacks that preserve these properties while manipulating the payload. Digital twins (DTs) have been used to emulate CAN traffic and generate attack scenarios for evaluating IDSs, but using a DT for intrusion detection remains unexplored. The objective of this study is to develop a DT-based IDS that identifies CAN attacks through residuals between predicted and observed powertrain signal behavior.

**Methods**: A shared-encoder LSTM DT was trained on 17 decoded signals from a real Hyundai/Kia CAN log, spanning engine and transmission messages, to jointly predict seven numeric signals and two categorical gear signals over a 24-step window. A timestep is flagged when a residual exceeds a calibrated threshold, and an adaptive rollout mechanism protects the twin's input history from sustained contamination. Four message-level attacks (plateau, continuous drift, masquerade, and gear masquerade) were evaluated against the twin and a range-and-plausibility baseline adapted from prior work.

**Findings**: The proposed DT outperformed the baseline across all attack types, achieving detection rates of 94.6% for continuous drift and 89.2% for masquerade attacks, while the baseline detected almost none of the fabricated payload attacks. These results demonstrate that learning coupled vehicle dynamics enables detection of stealthy payload manipulations that preserve normal CAN communication patterns. False positive rates reached 39.6%, highlighting the need for improved robustness under sustained attacks.

**Novelty**: This study introduces a DT-based CAN intrusion detection framework that learns the physical relationships among decoded vehicle signals and detects stealthy payload attacks through deviations between predicted and observed vehicle behavior rather than timing or protocol characteristics.

**Practical Applications:** The presented DT-based IDS represents a promising approach for detecting stealthy payload-level CAN attacks that preserve normal communication patterns, supporting behavior-based cybersecurity for connected and automated vehicles.

## INTRODUCTION

Automobiles increasingly rely on the Controller Area Network (CAN) to connect electronic control units (ECUs) governing braking, steering, powertrain, and other safety-critical functions. Introduced in the 1980s as a lightweight, real-time bus for connecting a handful of components, CAN now links dozens of ECUs in a typical passenger vehicle, yet its core design has changed little. Despite subsequent extensions to the CAN ecosystem, the legacy CAN protocol remains a broadcast communication system that provides no native support for message authentication, encryption, or sender verification. According to the legacy CAN protocol, any node that gains bus access, whether physically through a diagnostic port or remotely through a compromised infotainment or telematics unit, can inject or alter messages that every other ECU will accept as authentic. Hoppe and Dittman (2007) demonstrated the first documented CAN attack in a simulation environment, replaying captured messages to operate an electric window lift. Koscher et al. (2010) later gained physical access through a vehicle's OBD-II port and disabled a moving vehicle's engine and brakes while falsifying its instrument cluster display. Later, Valasek and Miller (2015) demonstrated a fully remote attack on Jeep Cherokee vehicles, requiring no physical access at all, that disabled braking and steering and ultimately triggered a 1.4-million-vehicle recall. These attacks demonstrate the threat posed by CAN intrusion attacks.

An intrusion detection system (IDS) is widely regarded as the most practical defensive measure once an attacker gains access to a vehicle's internal network. Existing automotive IDSs are generally divided into signature-based and anomaly-based approaches (Jin et al. 2021; Olufowobi et al. 2020; Halder et al. 2020; Wu et al. 2020). Signature-based methods define rules describing legitimate communication and known attack patterns, raising an alarm whenever observed network traffic violates these expectations. On a CAN bus, such rules are typically derived from message frequencies, identifier sequences, inter-frame timing, payload values, or combinations thereof. However, the heterogeneous and high-dimensional nature of CAN traffic, where numerous identifiers carrying different signals are transmitted at different rates, makes it difficult to formulate rules that generalize across operating conditions (Hanselmann et al. 2020). Because these systems ultimately rely on predefined signatures or behavioral constraints, they must be updated as new attack strategies emerge. As such, they often struggle to detect previously unseen or carefully crafted attacks (Wu et al. 2020). Moreover, sophisticated adversaries can preserve normal communication patterns by suppressing the legitimate transmitter through a bus-off attack and simultaneously transmitting forged messages under the victim ECU's identity (Cho and Shin 2016a; Bloom 2021). Other attacks exploit weaknesses at the CAN data-link layer to manipulate payloads while remaining protocol compliant (de Faveri Tron et al. 2022). Others exploit vulnerabilities arising from physical-layer implementation differences, such as ECU sampling-point configurations (Mohammed et al. 2022; Yue et al. 2021). These developments suggest that monitoring message timing or ordering alone is often insufficient to detect instrusions and requires further reasoning. For example, detecting advanced masquerade and payload-injection attacks ultimately requires reasoning about the message contents themselves.

Recent anomaly-based IDSs have therefore shifted their focus from communication patterns to the payload of CAN messages. One family of methods uses autoencoders to reconstruct normal behavior and identifies attacks through reconstruction error, whether reconstructing raw CAN traffic (Longari et al. 2021), decoded vehicle signals at multiple temporal scales (Shahriar, Xiao, et al. 2023), or manually defined groups of correlated signals (Novikova et al. 2020). Although effective, these approaches learn statistical regularities from the training data rather than explicitly modeling the physical relationships among vehicle ECU signals. A second family of anomaly-based IDSs formulates intrusion detection as a prediction problem, forecasting the next observation and comparing it with the actual message, either over raw identifier-payload sequences (Cobilean et al. 2023) or joint identifier-payload representations (Q. Liu et al. 2026). Since these methods operate directly on CAN communication rather than decoded physical quantities, they cannot explicitly exploit the underlying vehicle dynamics responsible for ECU signal payloads. Moriano, Bridges, and Iannacone (Moriano et al. 2022) move closer to accounting for vehicle dynamic behavior by detecting masquerade attacks through disruptions in the correlation structure among decoded signals, although these relationships are assessed through correlation similarity instead of being learned by a predictive model. Another line of work based on classical process monitoring, including PASAD (Aoudi et al. 2018), its CAN adaptation CASAD (Nowdehi et al. 2019), and the isolation-forest approach of Duan et al. (2023), identifies deviations from learned normal behavior in individual signals or payload-byte streams without modeling interactions among multiple signals. Despite their methodical differences, these approaches share a common limitation, in that none of them learn a predictive model of the coupled physical dynamics governing multiple vehicle ECU signals, motivating the digital-twin-based approach developed in this study.

Digital twins (DTs), first envisioned by Grieves in 2003 as virtual representations of physical systems throughout their lifecycle, have emerged as a powerful framework for monitoring, simulation, and prediction in cyber-physical systems. By continuously integrating observations from a physical asset with a corresponding virtual model, a DT can estimate the system's current state and predict its future behavior, enabling fault diagnosis, predictive maintenance, and increasingly, cybersecurity applications (Grieves and Vickers 2017; M. Liu et al. 2021). Rather than relying on predefined rules or signatures, security-oriented DTs detect anomalies by comparing observed system behavior against the behavior predicted by the virtual model, with deviations indicating potential faults or attacks (Eckhart et al. 2019; Dietz et al. 2020). This predictive paradigm is particularly attractive for CAN intrusion detection because the underlying vehicle dynamics impose strong physical relationships among signals that an attacker must preserve to remain undetected.

DTs have recently been extended to automotive CAN networks to emulate ECU interactions and generate realistic attack scenarios for IDS evaluation, demonstrating that ECU behavior and subsequent CAN traffic can be reproduced in a virtual or digital counterpart (Sharmin et al. 2025). However, this effort uses the DTs to evaluate IDSs rather than to perform intrusion detection itself. Whether a learned DT can directly identify stealthy CAN attacks by predicting the physically consistent evolution of decoded vehicle ECU signals remains largely unexplored. Thus, in this study, we present a DT-based IDS that learns the joint physical dynamics of different sensor values present in decoded powertrain signals and identifies attacks through mismatch between predicted and observed vehicle behavior. Rather than modeling message timing, protocol statistics, or individual signals in isolation, the DT developed in this study captures the physical coupling among related vehicle signals and uses prediction residuals to detect stealthy payload manipulations that preserve normal CAN traffic characteristics.

This paper develops a digital-twin-based intrusion detection framework that models the joint physical dynamics of decoded vehicle powertrain signals within a single predictive architecture, enabling the detection of stealthy payload manipulations through deviations between predicted and observed vehicle behavior. The developed framework is evaluated using real-world driving data against a diverse set of stealthy CAN attack scenarios, including both established and domain-specific attacks, and benchmarked against representative anomaly-detection baselines. Together, these results demonstrate the potential of physically grounded digital twins as a practical and effective framework for detecting sophisticated payload-level CAN attacks that preserve normal communication characteristics.

## LITERATURE REVIEW

Beginning with Hoppe and Dittman's (2007) initial demonstration on an electric window lift, subsequent research has confirmed the CAN bus vulnerabilities across a wide range of vehicle systems (Koscher et al. 2010), including remote exploitation of production vehicles without physical access (Woo et al. 2015; Valasek and Miller 2015). Hoppe et al. (2011) further classified CAN attacks into categories such as denial-of-service, fabrication, and unauthorized actuation. These demonstrations, ranging from simulated testbeds to production vehicles, established CAN security as a practical rather than theoretical concern before the defensive measures were developed.

Early detection methods exploited the CAN bus's protocol-level patterns rather than the payload of individual messages. Physical-layer fingerprinting authenticates the originating ECU of a message from small hardware variations rather than from message content: clock-skew-based fingerprinting estimates each ECU's intrinsic clock characteristics to detect sender impersonation with reported false-positive rates below one tenth of one percent (Cho and Shin 2016b), while voltage-based methods measure the differential signal each transceiver produces on the bus similarly (Cho and Shin 2017; Kneib and Huth 2018). Statistical methods instead model CAN stream-level regularities directly. Entropy-based detection monitors the information entropy of bus traffic to identify deviations from a learned baseline (Müter and Asaj 2011). A more recent work extends this idea with a more comprehensive collection of time-series and statistical features, and applies unsupervised anomaly detection to improve on sophisticated attacks (Shahriar, Lou, et al. 2023). ID-sequence similarity methods instead compare the order in which message identifiers are observed against a model of normal sequencing, using either basic pattern matching (Marchetti and Stabili 2017) or dynamic time warping to quantify how much an observed sequence deviates from typical traffic (Sun et al. 2022). Frequency-based detection, amongst the earliest approaches applied to this problem, flags CAN identifiers whose transmission frequencies deviate from their expected period (Taylor et al. 2015). Wu et al. (2020) provide a comprehensive survey of this research landscape covering rule-based, statistical, and early machine-learning-based defenses. Collectively, these methods are fast, lightweight, and interpretable, which makes them well-suited for resource-constrained in-vehicle deployment, but they are tailored to disruptions in message timing, frequency, or sequencing, and by design cannot detect an attack that preserves all three while only altering the message payload.

Machine learning and deep learning methods extend detection into the message payload, at the cost of greater computational overhead and, in most cases, a need for representative training data across a vehicle's full operating range. Sequence models based on long short-term memory (LSTM) networks have been applied both to raw CAN traffic (Hossain et al. 2020; Taylor et al. 2016) and to the byte-level payload directly (Kang and Kang 2016), while a related specification-based approach models the timing behavior of individual identifiers as an anomaly-detection problem rather than a purely statistical one (Olufowobi et al. 2020). Convolutional architectures adapted from image classification have been applied to CAN frames reshaped as image-like matrices (Song et al. 2020), and generative adversarial networks have been trained on normal CAN traffic and used for detecting unknown attack types (Seo et al. 2018). More recently, transformer-based attention networks have been applied to the same sequence-classification problem, leveraging self-attention to model long-range dependencies within CAN message sequences (Nguyen et al. 2023). Graph-based architecture represents CAN identifiers as nodes connected by transmission relationships. One approach uses threshold classifiers over graph features (Park et al. 2023) to detect anomalies, while another employs graph attention networks that learn which relationships matter most for a given attack (Xiao et al. 2024). Hybrid and ensemble variants that combine several of these architectures (Al-Aql 2024; Javed et al. 2021), or optimize their hyperparameters directly, have also been proposed (Ileri et al. 2025). Almost all of this literature, however, is evaluated against fabrication-style benchmarks such as Car-Hacking and OTIDS, in which injected messages disturb per-ID timing statistics that a representative model can learn to recognize.

A different class of techniques adopts an explicitly predictive or reconstructive framing that does not rely on timing disruption at all. CANnolo (Longari et al. 2021) trains a single LSTM autoencoder on normal traffic and reconstructs the CAN traffic sequence, detecting anomalies from the reconstruction error. CANShield (Shahriar, Xiao, et al. 2023) instead trains an ensemble of convolutional autoencoders, each reshaping a sliding queue of signal values into an image-like matrix at a different temporal scale, and combines their reconstruction losses into a single ensemble score. It shares our twin's residual-based detection philosophy, and it focuses on decoded signal values rather than raw payload bytes. But this method learns statistical relationships between signals, whereas our approach learns predictive physical relationships among signals. Moriano, Bridges, and Iannacone (Moriano et al. 2022) target masquerade attacks more directly, hypothesizing that such attacks disrupt the correlation structure among signals. It detects them by comparing the clustering similarity of signal correlations in a test window against a clean baseline. The premise closely aligns with our twin's own use of cross-signal physical relationships, though it is implemented as a correlation-similarity check in (Moriano et al. 2022) rather than a learned predictive model. Novikova et al. (Novikova et al. 2020) studies a related idea by manually grouping highly correlated signals into subgroups and training a separate autoencoder per subgroup. Our approach differs from it by grouping correlated and physically related signals into a single encoder to represent the system's state. Cobilean et al. (Cobilean et al. 2023) predict the next CAN identifier and payload sequence with a transformer and compare the prediction against what is actually observed. It is similar to the predict-and-compare logic used in this paper, except that it is applied here to raw CAN sequences rather than decoded physical signals. Most recently, MIDS (Q. Liu et al. 2026) targets the masquerade setting directly, using a bidirectional state-space model to reconstruct joint identifier-payload semantics and reporting strong generalization across several public benchmarks. Duan et al. (Duan et al. 2023) address the related problem of data-tampering detection directly, using an improved isolation-forest method sensitive to local anomalies in the data field rather than to the message stream as a whole. Outside the deep-learning literature, PASAD (Aoudi et al. 2018) introduced a departure-based anomaly detection framework for industrial control systems using singular spectrum analysis. CASAD (Nowdehi et al. 2019) subsequently adapted this methodology to CAN payload-byte streams, detecting departures from learned normal behavior without requiring knowledge of the encoded vehicle signals. Unlike these systems, our twin-based approach learns the physical coupling between decoded vehicle signals within a single predictive model. It combines the cross-signal reasoning exemplified by Moriano et al. with the learned autoregressive prediction used by the autoencoder- and transformer-based methods, while extending the departure-based philosophy of PASAD and CASAD from single process or payload-byte streams to a shared model of physically meaningful vehicle signals.

Collectively, these studies demonstrate the effectiveness of reconstruction-, prediction-, and departure-based anomaly detection, but none explicitly models the joint physical dynamics governing multiple decoded vehicle signals. Our DT-based approach this gap by learning these relationships directly and detecting attacks through departures from physically consistent vehicle behavior

## METHODS

The intrusion detection framework we present in this study is built around a DT that comprises a shared-encoder LSTM trained on attack-free driving data to learn the combined dynamics of a vehicle's powertrain and transmission control subsystems. At each timestep, the twin estimates the expected value of each monitored signal from a fixed-length historical window, and the detection process is based on a per-timestep comparison of this predicted value against the observed CAN value. Whenever the residual exceeds a calibrated threshold or, for categorical signals, when a misclassification occurs, a violation is flagged. A second mechanism uses the same per-timestep violations to protect the twin's own future predictions from degrading under a sustained attack. The detection system tracks each monitored signal independently to determine whether its recent violation rate has exceeded a specific threshold. If it does so, the twin's own prediction replaces that signal's observed value. This mechanism, called adaptive rollout, prevents a continuous attack from being fed into the model's context and gradually steering its own future predictions toward the attacker's manipulated values. A more conservative condition determines the restoration of normal operation, and a periodic anchoring step forces trust in observed CAN values at fixed intervals, irrespective of the current state, thereby bounding how far this substitution can drift over time. **Figure 1** summarizes this workflow. This framework is evaluated against four attack scenarios applied to driving data, selected to represent distinct real-world spoofing strategies, and compared against a range and plausibility-based anomaly detector (Müter et al. 2010).

### Dataset, Signal Decoding, and Signal Selection

The dataset considered in this study is a real-world Hyundai/Kia CAN log (Seo et al. 2018) decoded using the Python *cantools* library, based on two DBC files, namely *hyundai_kia_generic.dbc* for engine-management messages and *hyundai_santafe_2007.dbc* for transmission control messages. From the entire CAN log four messages were chosen, namely *EMS11 (0x316)*, *EMS12 (0x329)*, and *EMS14 (0x545)* for engine management and *TCU_Data (0x43F)* for transmission control. All four messages are broadcast independently and transmitted at different rates, so a shared time base was required before they could be used jointly. Each message was decoded, sorted by timestamp, and aligned onto a shared 50-ms grid using a *merge_asof* operation such that each grid point takes the most recently observed value for each signal, avoiding any look-ahead. A forward/backward fill was also applied to handle boundary cases at the start and end of the sequence. The 50-ms interval was selected based on the representative update interval observed across the full set of CAN traffic during initial exploratory analysis. Before model development, a correlation analysis was conducted across the ten numeric physically relevant signals *(N, TQI, TQI_ACOR, TQFR, VS, TPS, TEMP_ENG, VB, BAT_Alt_FR_Duty, InputShaftSpeed*) present in these four messages from an attack-free driving session. A Pearson correlation analysis was conducted to assess the relation between these signals, followed by a lagged cross-correlation analysis to characterize whether any signal pairs exhibited a meaningful lagged relationship. This analysis established the physical coupling that justified modeling these signals jointly. *N* correlates strongly with *InputShaftSpeed* (Pearson r = 0.83), and more moderately with *TPS* (Pearson r = 0.68), *TQI* (Pearson r = 0.61), and *VS* (Pearson r = 0.53). *TQI* correlates strongly with *TPS* (Pearson r = 0.88), being the strongest relationship in the signal set, while *VS* correlates strongly with *InputShaftSpeed* (Pearson r = 0.77). Together, *N*, *TQI*, *TPS*, *VS*, and *InputShaftSpeed* form a tightly coupled physical subsystem, representing the vehicle engine and transmission system jointly. Some signals displayed weak correlation. For example, *VB* and *BAT_Alt_FR_Duty* both showed weak correlation with every other signal. During training, *BAT_Alt_FR_Duty* exhibited relay-like behavior that the encoder could not learn reliably. Its residual was thus excluded from the anomaly detection mechanism, though it was kept as a regression target for training. In contrast, *VB*'s weak correlation is due to the low signal variance. These signals were kept as they represent distinct, modest information regarding the subsystem's state. *TQI_ACOR* was found near-redundant with *TQI* (Pearson r = 0.999) and was dropped. *TQFR* showed a strong, physically meaningful correlation not only with *N* (Pearson r ≈ −0.82) but with several other core drivetrain signals already in the model, including *InputShaftSpeed* (Pearson r ≈ −0.78) and *VS* (Pearson r ≈ −0.67). This indicated that its behavior was largely redundant with information the shared encoder already captures rather than reflecting an independent physical relationship. *TEMP_ENG* showed no fast changes that kept pace with any other signal and was retained only as a conditioning input, never as a prediction target. In *TCU_Data*, three categorical signals were also present: *CurrentGear*, *GearSwitch*, and *SelectorPosition*. Of these, *SelectorPosition* was dropped as it was essentially a redundant proxy for *CurrentGear*, while *CurrentGear* and *GearSwitch* were both retained as classification targets. *CurrentGear* and *GearSwitch* were both one-hot encoded to generate 7 and 2 unique values, respectively. This brought the total inputs to 17, including 8 numeric inputs, 2 encoded *GearSwitch* inputs and 7 encoded *CurrentGear* inputs. **Table 1** summarizes the final signal set.

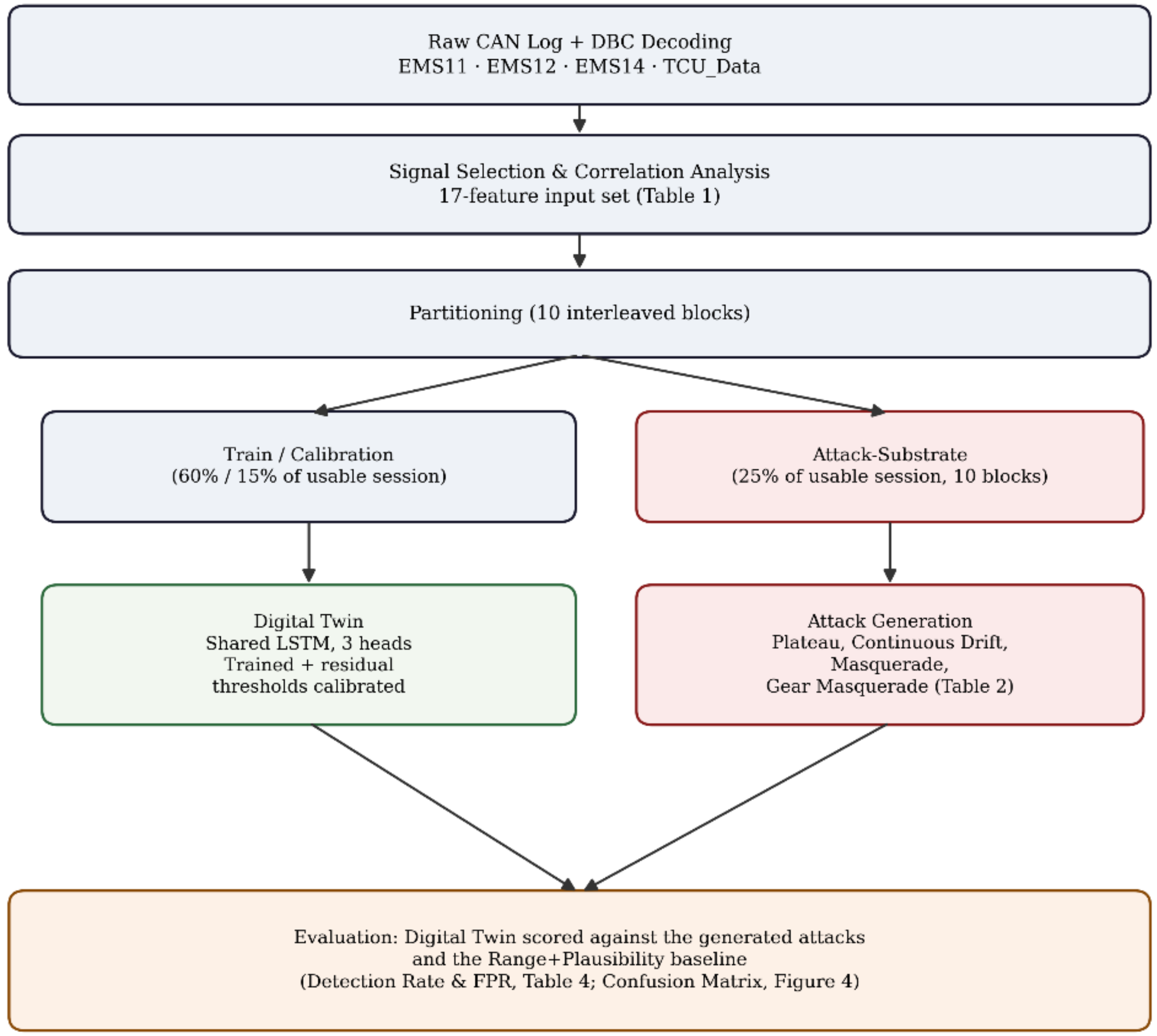

**Figure 1 DT-based CAN IDS workflow**

**TABLE 1 Final CAN Signal Set**

| Signal | Data Type | Role | Source Message |
|---|---|---|---|
| **N** | numeric | input + regression target | *EMS11 (0x316)* |
| **TQI** | numeric | input + regression target | *EMS11 (0x316)* |
| **TQI_ACOR** | numeric | excluded from twin due to redundancy | *EMS11 (0x316)* |
| **TQFR** | numeric | excluded from twin due to redundancy | *EMS11 (0x316)* |
| **VS** | numeric | input + regression target | *EMS11 (0x316)* |
| **TPS** | numeric | input + regression target | *EMS12 (0x329)* |
| **TEMP_ENG** | numeric | input only (conditioning signal; not a regression target) | *EMS12 (0x329)* |
| **VB** | numeric | input + regression target | *EMS14 (0x545)* |
| **BAT_Alt_FR_Duty** | numeric | input + regression target (residual excluded from detection gating) | *EMS14 (0x545)* |
| **InputShaftSpeed** | numeric | input + regression target | *TCU_Data (0x43F)* |
| **CurrentGear** | categorical (7 classes) | one-hot input + classification target | *TCU_Data (0x43F)* |
| **GearSwitch** | categorical (2 classes) | one-hot input + classification target | *TCU_Data (0x43F)* |
| **SelectorPosition** | categorical (2 classes) | excluded (redundant proxy for CurrentGear) | *TCU_Data (0x43F)* |

**Partitioning**

The raw log spans an extended vehicle operation session that includes a period of pre-drive idling before the vehicle is put into gear, and a reverse maneuver after the drive concludes. The usable driving session was therefore defined as the span from the first sustained engagement of gear 1 (marking the start of active driving) to the first sustained engagement of the reverse gear occurring after that point (marking the end of the drive), yielding 7,777 usable rows. This session was then divided into 10 adjacent blocks and partitioned 60/15/25 into train, calibration, and attack-substrate sets, with each partition drawing proportionally from across all blocks rather than from one contiguous span, avoiding bias toward whichever driving conditions happened to occur early or late in the session. The resulting partition sizes are: 4,660 rows for training, 1,160 rows for calibration, and 1,957 rows for attack substrate. Normalization statistics (mean and standard deviation per numeric signal) were computed exclusively on the train partition and applied to calibration and attack-substrate data, preventing any leakage of reserved statistics into model fitting.

### DT Architecture

The twin here is a single shared LSTM (hidden layer size = 128) operating on a 24-step (1.2 s) input window over 17 features, with three output heads: numeric regression over seven targets (excluding *TEMP_ENG*), CurrentGear classification, and GearSwitch classification. An exploratory lagged cross-correlation analysis across the candidate signal set found no reliable inter-signal delay exceeding a couple of timesteps at the 50-ms sampling rate. The 24-step window was therefore adopted as a generous margin, comfortably exceeding any measured short-timescale coupling while still spanning the duration of a typical gear change or throttle transition. The model is trained exclusively on the train partition, using the normalization statistics defined in the previous section. Training used the Adam optimizer with a learning rate of 1e-3 and a batch size of 64, run for 50 epochs. The numeric regression head is trained with Huber loss, chosen for its reduced sensitivity to occasional outlier readings relative to squared error, while the CurrentGear head uses standard cross-entropy, and the GearSwitch head uses class-weighted cross-entropy to counter the inherent rarity of shift events. All three losses are summed with equal weighting into a single training objective as shown in **Equation 1**. The checkpoint used for evaluation is the epoch with the lowest total loss on the calibration partition, rather than the final epoch's weights, to guard against evaluating an overfit late-stage model.

$$L = L_{Huber(regression)} + L_{Cross\ entropy(CurrentGear)} + L_{Cross\ entropy,weighted(GearSwitch)} \quad (1)$$

## EVALUATION

To evaluate our DT, we designed three attack models based on prior literature. We added a detection logic to our DT and modified it to handle different attack types. We also included a real-time test using CAN hardware to test its real-life feasibility. The following sections discuss each in detail.

### Attack Models

Attack scenarios were designed to align with the synthetic attack taxonomy introduced by the SynCAN dataset (Hanselmann et al. 2020), which defines several canonical signal-level attack types for CAN intrusion detection evaluation. All attacks are message-level substitutions. The attacker is assumed to have silenced the legitimate transmitting ECU, a capability separately demonstrated in prior work on CAN bus-off attacks (Bloom 2021), and to transmit fabricated values in its place. Three attack types were constructed on the Attack-substrate partition (1,957 rows, 25 percent of the usable session, held out from both training and calibration): plateau, in which a signal is frozen at a fixed value for a sustained duration; continuous, in which a signal's value is progressively displaced from its true value at a constant per-step rate; and playback, in which a signal's true values are replaced with genuine values drawn from an earlier segment of the same block, resulting in a masquerade-style replay attack. These three attacks target *N*, *TQI* and *VS*, the three signals carried in a single engine message, *EMS11* (*0x316*). The continuous attack's per-step drift rate is set to the mean absolute step-to-step change of each signal, computed over the train partition rather than the attack data itself, reflecting a rate an attacker could plausibly estimate from historical logs. Attack duration was set at 40 percent of each block's length (approximately 78 steps, or 3.9 s at the 50-ms grid), beginning at the midpoint of each block; this was chosen to sit just under the 4-s upper bound of SynCAN's documented 2–4 s (Hanselmann et al. 2020) attack-interval range, keeping attack duration comparable to that prior benchmark while scaling naturally with block length. A fourth attack, targeting the gear-switching signals, was additionally constructed to probe the discrete/classification side of the twin. This attack is specific to the gear signals evaluated here and falls outside SynCAN's original continuous-signal-only taxonomy. It applies the same playback/masquerade construction described above, but to *InputShaftSpeed*, *CurrentGear*, and *GearSwitch*, all three carried in the transmission message *TCU_Data* (*0x43F*). This attack substitutes a genuine earlier segment of the block into all three signals simultaneously from the same donor window, preserving physical coherence among the gear-cluster signals while remaining inconsistent with the vehicle's actual state. All four attacks were evaluated across all ten attack-substrate blocks. **Table 2** summarizes all four attack scenarios.

**TABLE 2 Attack Scenarios**

| **Attack** | **Target Signal(s)** | **Source Message** | **Construction** |
|---|---|---|---|
| **Plateau** | *N, TQI, VS* | *EMS11 (0x316)* | Frozen at a fixed value for 40% of block length |
| **Continuous (drift)** | *N, TQI, VS* | *EMS11 (0x316)* | Displaced at a constant per-timestep rate (mean absolute step-to-step change from train partition) from the last correct value for 40% of block length |
| **Playback (masquerade)** | *N, TQI, VS* | *EMS11 (0x316)* | Replaced with genuine values from an earlier segment of the same block, for 40% of block length |
| **Gear-cluster masquerade** | *InputShaftSpeed, CurrentGear, GearSwitch* | *TCU_Data (0x43F)* | Replaced with genuine values from an earlier segment of the same block, for 40% of block length |

**Range and Plausibility Baseline**

As a lightweight, non-learned baseline, we implement two of the eight anomaly detection sensors proposed by Müter, Groll, and Freiling (Müter et al. 2010), namely, the Range Sensor and the Plausibility Sensor. The Range Sensor flags a signal whenever its decoded value falls outside the bounds specified in the vehicle's own DBC signal definitions (e.g., *N: 0-16383.75 rpm*, *VS: 0-254 km/h*), following the study's classification of the sensor being specification-based. The Plausibility Sensor flags a signal whenever its per-step change exceeds the largest step-to-step change observed for that signal on the train partition. This adapts the study's plausibility check to a setting without documented physical rate limits. Both sensors operate per-signal, matching the twin's own monitoring scope. The remaining six sensors described in the original paper (Müter et al. 2010) were not implemented. Formality and Frequency are inapplicable by construction, since our attack model preserves valid frame structure and message timing. Location, Correlation, Protocol, and Consistency require multiple bus systems or genuinely redundant sensor sources not present in our dataset. Both sensors evaluate each newly observed sample independently, keeping the comparison directly aligned with the twin's own detection metric.

**Attack Detection and Adaptive Rollout Mechanisms**

The IDS analyses signals on a per-timestep basis. At each timestep, a signal's residual (predicted minus observed) is compared against a signal-specific threshold. Residuals are determined according to **Equation 2,** $\hat{y}_t$ is the predicted output and $y_t$ is the real observed value. Residual thresholds are calibrated on the calibration partition, separate from both training and attack construction. For each of the six detection-eligible numeric signals (all numeric regression targets except *BAT_Alt_FR_Duty*), a timestep is flagged if the absolute residual *K* exceeds *3* times the standard deviation of the absolute residual observed on calibration data for that signal. *CurrentGear* and *GearSwitch* however use a fixed misclassification threshold. For these two categorical signals, a timestep is flagged whenever the predicted class differs from the observed class.

$r_t = |\hat{y}_t - y_t|$ (2)

A separate mechanism, adaptive rollout, protects the twin's own autoregressive input window. Because the twin conditions each predict on its own recent input history, a sustained attack risks contaminating that history and degrading the twin's subsequent predictions. Adaptive rollout is a per-signal passive/closed-loop state machine. If the rate of spot violations within a trailing window (20 timesteps) exceeds a reject-rate threshold (60%), the affected signal enters closed-loop mode, in which the twin substitutes its own prior predictions for that signal rather than the observed values. The signal returns to passive mode once the violation rate over a trailing reacquire window (40 timesteps) falls to or below a reacquire-rate ceiling (10%). For any signal currently in closed-loop mode, this override is periodically suspended: a shared, time-based anchor window re-admits raw sensor data for 10 timesteps out of every 200, independent of which signal is in closed-loop or whether an attack is active at that moment. This prevents a closed-loop signal from drifting indefinitely from the actual vehicle state, at the cost of occasionally reintroducing corrupted data. The adaptive rollout parameters are given in **Table 3**, and they were determined by preliminary experimentation and held constant across all scenarios.

**Real-Time Hardware Testbed**

To evaluate the feasibility of the framework in a more realistic setting, the twin-based IDS was deployed on a small-scale hardware-in-the-loop testbed. Arduino nodes were configured to broadcast the *EMS11*, *EMS12*, *EMS14*, and *TCU_Data* messages from the attack substrate partition at their original recorded timing, reproducing the same stream content, message timing, and frequency as in the offline test. The Arduinos were connected with MCP2515 interfaces and configured to send CAN messages over a CAN bus, using the *mcp_can* library. One Arduino node was assigned for sending the 3 *EMS* messages, reflecting the engine control ECU and one node was assigned for sending the *TCU_Data* messages reflecting the transmission control ECU. A third Arduino node was configured to represent the attacker node that sends the fabricated messages. The legitimate engine and transmission nodes were configured to stop broadcasting over the duration of the attacks, reflecting that the attacker had taken control of them. A desktop computer running the twin-based IDS was connected to the CAN bus through a USB-CAN FD interface. The detection was performed by running the IDS in real-time. Incoming messages were decoded and aligned to the 50-ms grid and passed into the 24-step input window as they arrived. The same detection procedure used in the offline version was also used here. We evaluated our IDS in this real-time setup using one representative block from each of the four attack scenarios.

**TABLE 3 Adaptive Rollout Parameters**

| **Parameter** | **Value** |
|---|---|
| Reject window | 20 |
| Reject rate | 60% |
| Reacquire window | 40 |
| Reacquire rate | 10% |
| Anchor interval | 200 |
| Anchor duration | 10 |

**Evaluation Metrics**

To evaluate the performance of our IDS, we chose several metrics to determine its effectiveness in both offline and real-time configurations. Detection rate and false positive rate were the primary metrics chosen to evaluate the IDS's offline performance across the ten blocks. They are defined in **Equation 3** and **Equation 4**:

$$Detection\ Rate = \frac{True\ Positive}{True\ Positive + False\ Negative} \quad (3)$$

$$False\ Positive\ Rate = \frac{False\ Positive}{False\ Positive + True\ Negative} \quad (4)$$

Because block-to-block variance in these metrics is substantial, results are reported as *mean ± standard deviation* across the 10 attack-substrate blocks rather than as bare means. Additionally, for the hardware-in-the-loop test, several other metrics were selected to evaluate the feasibility of this IDS in real-time attack conditions. Three metrics were chosen for this purpose, namely, detection latency and per-message processing time. Detection latency is the time between a spoofed message being received across the network and the IDS flagging it as spoofed. It measures the time between the attack taking place and the alarm being sounded.

## RESULTS AND DISCUSSION

**Table 4** presents the aggregate performance of our twin-based approach and the compared baseline on all 10 blocks across the 4 attacks scenarios. The DT-based anomaly detector presented in this study outperforms the baseline by a wide margin across all of the evaluated attack types. The baseline's poor performance can be attributed to the fact that all four attacks were crafted to ensure that the fabricated values were within the typical range observed in a CAN stream. Thus, all of the attack scenario data was likely within the typical range and the change between two steps was also likely within the normal range. In the case of our twin-based detector, we can observe that it performs best on the drift attack, followed by masquerade and plateau, with the gear masquerade performance being the worst. The standard deviations indicate that our twin-based detector can reliably detect the drift attack, but consistency drops for the masquerade attack. In the case of plateau and gear masquerade, the high standard deviations indicate uneven detector performance across blocks and that some blocks exhibiting poor performance. In the case of the drift attack, the signals likely diverge in a way that their internal relations are no longer maintained. Consequently, the twin detects it reliably and accurately.

**Figure 2** presents the plots for the continuous drift and plateau attacks, while **Figure 3** presents the plots for the masquerade and gear masquerade attacks, each showing the real, attacked, and twin-predicted trajectories for a representative attack-substrate block. **Figure 4** presents the aggregate confusion matrices for the twin-based detector, summarizing the block-averaged true positive, false positive, false negative, and true negative counts across all four attack types.

In **Figure 3**, with regards to masquerade attacks, the twin's performance does not drop too much, but the reliability drops substantially. This is likely due to the masquerade attack behaving more like a real signal and, in some cases, coming close to the attack-free values. We can see this in the plot in **Figure 3**, where the masquerade intersects the attack-free signal and the detections drop in that region. **Figure 4** also demonstrates that the twin-based detection has a similar aggregate performance for drift and masquerade. In the case of the plateau, the drop in detection rate is likely due to the fact that the CAN signal streams feature periods when the signal value is relatively constant. These attack-free near-constant stretches can be observed in the plots in **Figure 2** and **Figure 3**. The twin is inevitably fed some corrupt data at the attack onset. Even if adaptive rollout is triggered, the initial corrupt data can pull the twin's state into one where the signals are relatively constant. The twin can erroneously assume this to be a legitimate near-constant stretch as opposed to a plateau attack. This can be observed as well in **Figure 4**, where the twin is nearly as good at detecting negatives for plateau, as it is with masquerade and drift, but falls short when it comes to positives. **Figure 2** also displays some cases where the plateau attack is close to the attack-free signal or intersects it. The gear masquerade attack shows the worst performance, with the lowest average detection rate and a standard deviation close to it. The poor performance of the twin on the gear attack can be attributed to the fact that the vehicle typically maintains the same gear over long stretches of time. This led to the effect that a masquerade attack splice was sometimes identical to the normal data it was replacing. **Figure 2** demonstrates this effect for gear 2. One of the observations that stands out is the relatively high false positive rate across the board, ranging from 17.5% to 39.6%. It must be stressed that the twin is not a perfect counterpart of real vehicle subsystems, and it can erroneously flag a normal CAN payload if the residual is too high, which requires further investigation. Additionally, the twin tends to remain in closed-loop mode even after passing the attack region. This can be observed in **Figures 2** and **3**. The adaptive rollout is designed such that regaining trust in a signal is more difficult than losing it. This leads to the unfortunate case of the twin having a high overall FPR.

**Table 5** presents the real-time performance of our IDS on our CAN testbed. It can be observed that the mean detection latency is below one millisecond. The standard deviation indicates that the detection latency varies substantially between messages. The maximum detection latency across all attacks is 4.5-ms, which is below the lowest time interval (5.046-ms) observed for these messages (*EMS11, EMS12, EMS14* and *TCU_Data*). This confirms that the IDS is computationally fast enough to catch any spoofed frame. The mean model inference time is also below 1 millisecond. Looking at the maximum model inference time observed, we can see that it rises up to 4.40 ms. This value is still below the 50-ms time grid we established before. This indicates that the model is unlikely to be overflooded with spoofed messages. The model inference time suggests that the IDS can safely operate on a much data rate than the one used, and one that is much closer to the lower boundary of time intervals of these messages.

**TABLE 4 Detection Performance by Attack Types**

| Attack type | Average | Twin FPR % | Baseline Det. % | Baseline FPR % |
|---|---|---|---|---|
| **Plateau** | 61.8 ± 36.3 | 33.0 ± 27.1 | 0.0 ± 0.0 | 0.1 ± 0.3 |
| **Continuous (drift)** | 94.6 ± 4.8 | 39.6 ± 24.8 | 0.0 ± 0.0 | 0.3 ± 0.4 |
| **Masquerade** | 89.2 ± 21.5 | 34.5 ± 28.3 | 0.0 ± 0.0 | 0.1 ± 0.3 |
| **Gear masquerade** | 44.7 ± 37.3 | 17.5 ± 16.7 | 0.0 ± 0.0 | 0.1 ± 0.3 |

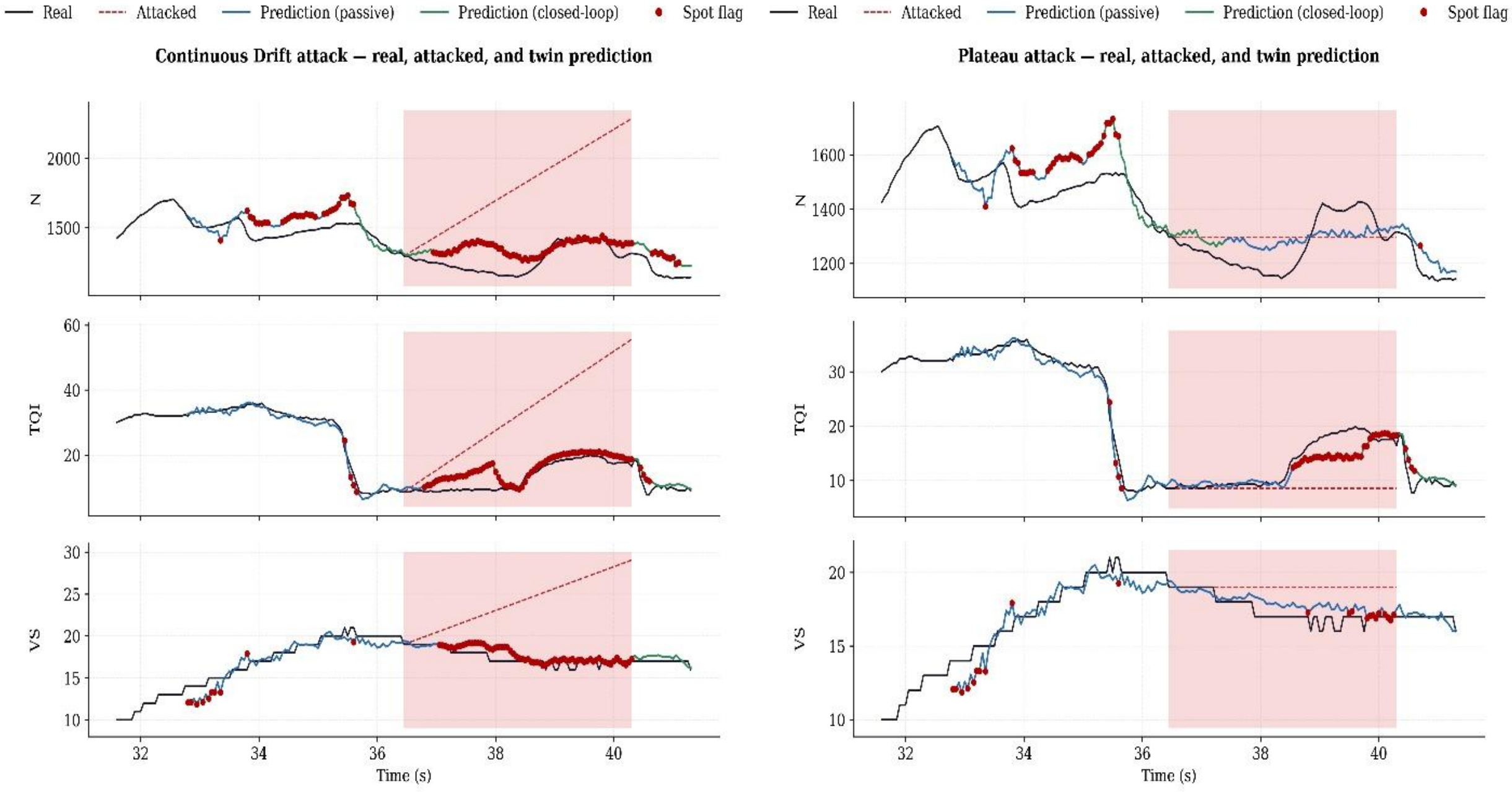

**Figure 2 Continuous drift (left) and plateau (right) attacks: real, attacked, and twin-predicted trajectories for a representative block**

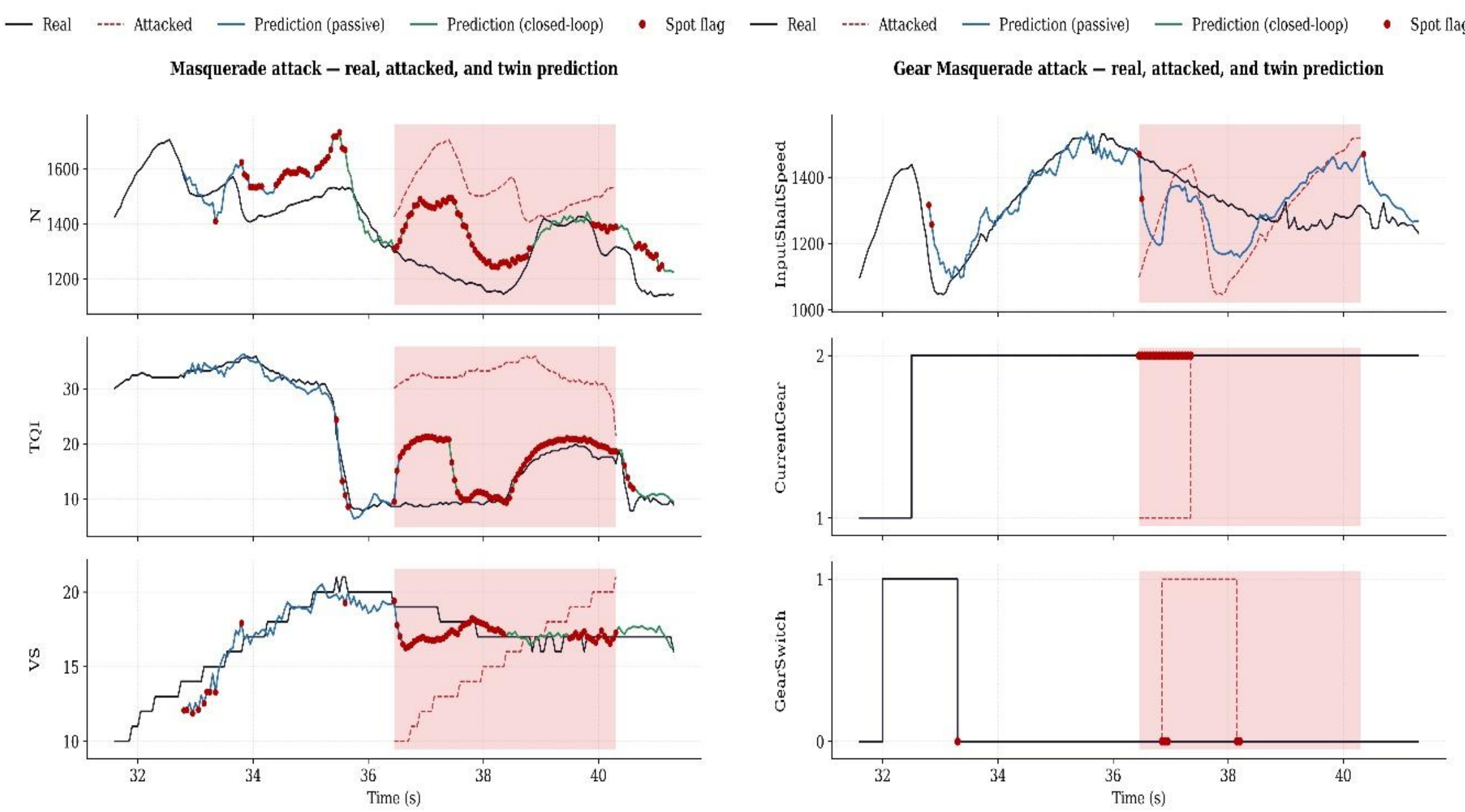

**Figure 3 Masquerade (left) and gear masquerade (right) attacks: real, attacked, and twin-predicted trajectories for a representative block.**

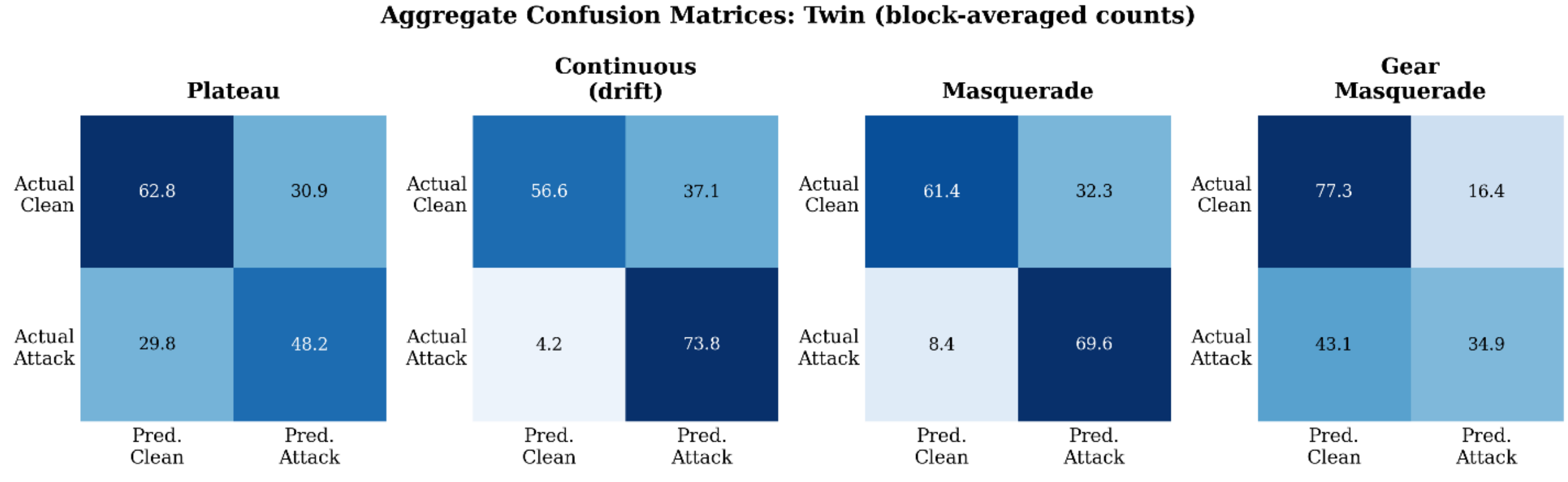

**Figure 4 Aggregate confusion matrices for the twin-based detector, block-averaged counts across all four attack types**

**TABLE 5 Real-time IDS Performance by Attack Types**

| Attack type | Average Detection Latency ± Standard Deviation (ms) | Maximum Detection Latency (ms) | Average Model Inference Time ± Standard Deviation (ms) | Maximum Model Inference Time (ms) |
|---|---|---|---|---|
| **Plateau** | 0.86 ± 0.59 | 3.82 | 0.78 ± 0.57 | 3.72 |
| **Continuous (drift)** | 0.68 ± 0.52 | 4.57 | 0.61 ± 0.46 | 4.40 |
| **Masquerade** | 0.68 ± 0.43 | 3.41 | 0.61 ± 0.42 | 3.21 |
| **Gear masquerade** | 0.81 ± 0.49 | 2.88 | 0.68 ± 0.40 | 2.72 |

## CONCLUSIONS

This study presents a DT-based IDS for vehicle powertrain CAN networks that detects attacks by comparing predicted vehicle behavior against observed CAN signals. Trained using real-world Hyundai/Kia driving data, the developed digital twin consistently outperformed baseline methods across the evaluated attack scenarios, with particularly strong performance against continuous-drift and masquerade attacks that violate learned physical relationships among powertrain signals. Detection performance was less consistent for plateau and gear-masquerade attacks, where manipulated values remained locally plausible despite being malicious, highlighting an inherent challenge for prediction-based anomaly detectors operating on physically constrained systems. These results highlight an inherent challenge for prediction-based detectors operating in physically constrained cyber-physical systems.

Overall, the findings demonstrate that modeling the coupled dynamics among decoded vehicle signals provides an effective approach for detecting stealthy payload manipulation without relying on message timing characteristics or protocol-level features. Future work will extend the digital twin presented in this paper to additional vehicle subsystems and investigate advanced temporal modeling approaches to improve robustness against attacks that closely mimic nominal vehicle behavior. In addition, future research will focus on generalizing DT-based IDS approaches to a broader range of sophisticated and emerging cyber attack scenarios.

## ACKNOWLEDGMENTS

The authors used Claude and ChatGPT to assist with editing portions of the manuscript, including advanced language editing and to assist with coding. The authors take full responsibility for the technical content, analyses, and conclusions presented in this paper

## AUTHOR CONTRIBUTIONS

The authors confirm contribution to the paper as follows: study conception and design: A. Rahman, M.S. Salek, M. Chowdhury; data collection: A. Rahman; analysis and interpretation of results: A. Rahman; draft manuscript preparation: A. Rahman, M.S. Salek, M. Chowdhury,. All authors reviewed the results and approved the final version of the manuscript.

## DECLARATION OF CONFLICTING INTERESTS

The authors declared no potential conflicts of interest with respect to the research, authorship, and/or publication of this article.

## FUNDING

This research was funded by the National Center for Transportation Cybersecurity and Resiliency (TraCR) (a US Department of Transportation National University Transportation Center) headquartered at Clemson University, Clemson, South Carolina, USA, under Grants: 69A3552344812, 69A3552348317.